%% file: paper.tex
\documentclass[reprint, aps,prd, preprintnumbers,groupedaddress,nofootinbib]{revtex4-1}

\pdfoutput=1
\usepackage{graphicx}
\usepackage{latexsym}
\usepackage{amsfonts}
\usepackage{amssymb}
\usepackage{amsmath}
\usepackage{slashed}
\usepackage{feynmp}
\usepackage{hyperref}
\usepackage{url}
\usepackage{color}
\usepackage{cancel}
\usepackage[normalem]{ulem}
\usepackage{multirow,array}

\usepackage{dcolumn}
\usepackage{bm}

\input{declare}

\begin{document}

\title{Boosting the $H\to$~invisibles searches with $Z$ boson polarization}

\author{Dorival Gon\c{c}alves} 
\email{dorival@pitt.edu}
\affiliation{PITT PACC, Department of Physics and Astronomy, University of Pittsburgh, 3941 O'Hara St., Pittsburgh, PA 15260, USA}
\author{Junya Nakamura}
\email{junya.nakamura@itp.uni-tuebingen.de}
\affiliation{Institut f\"ur Theoretische Physik, Universit\"at T\"ubingen, 72076 T\"ubingen, Germany}

\preprint{PITT-PACC-1814}

%%%%%%%%%%%%%%%%%%%%%%%%%%%%%%%%%%%%%%%%%%%%%%%%%%
\begin{abstract}

It is argued that, in $H \to $~invisibles searches with $Z(\ell\ell)H$ associated production at the LHC, the signal efficiency  can be sensibly improved
via a detailed study of the  $Z$ boson polarization,  discriminating  between the signal and the dominant-irreducible $Z(\ell\ell)Z(\nu\nu)$ background. 
We first present a comprehensive polarization study, obtaining the complete set of angular coefficients $A_i$  in the Collins-Soper reference frame and 
identifying the dominant phenomenological effects. Then, we show the results for a realistic Monte Carlo study to $H\to$~invisibles, taking the polarization 
analysis into account.  We obtain about $20\%$ improvement in the upper bound for  the branching ratio of the Higgs boson to
invisible particles, assuming  $300\ \mathrm{fb}^{-1}$ of data at  the 13 TeV LHC.

\end{abstract}

\maketitle

%%%%%%%%%%%%%%%%%%%%%%%%%%%%%%%%%%%%%%%%%%%%%%%%%%
\section{Introduction}

Bounding the invisible decay rate of the observed Higgs boson is one of the major targets of the LHC programme~\cite{Aad:2014iia, 
Chatrchyan:2014tja,  Khachatryan:2014jba, Aad:2015uga, Khachatryan:2015vta, Aad:2015txa,Aad:2015pla, Khachatryan:2016whc, 
Aaboud:2017bja,Choudhury:1993hv,Eboli:2000ze,Corbett:2015ksa}. While the  Standard Model (SM) predicts a very small rate 
$\mathcal{BR}_{H\to\mathrm{inv}} \simeq  0.1\%$~\cite{Heinemeyer:2013tqa}, there are many extensions of the SM,  refereed  to 
as Higgs Portal models~\cite{Patt:2006fw,Bento:2000ah,Englert:2011yb,Ipek:2014gua,Goncalves:2016iyg},  that predict a significantly 
larger $\mathcal{BR}_{H\to\mathrm{inv}}$. Therefore, the observation of invisible Higgs boson decays  above the small SM rate would
be a smoking gun  signature for physics beyond the SM and could be the first direct evidence for the underlying microphysics of the Dark Sector.

Direct searches for invisible decays of the Higgs boson have been actively conducted at the large hadron collider (LHC)
by the ATLAS and CMS Collaborations in several Higgs production channels~\cite{Aad:2014iia, Chatrchyan:2014tja,  Khachatryan:2014jba,
 Aad:2015uga, Khachatryan:2015vta, Aad:2015txa,Aad:2015pla, Khachatryan:2016whc, Aaboud:2017bja}.
From combinations of these searches, the current upper bounds are  $\mathcal{BR}_{H\to\mathrm{inv}} < 25\%$ by 
ATLAS~\cite{Aad:2015pla} and $\mathcal{BR}_{H\to\mathrm{inv}} < 24\%$ by CMS~\cite{Khachatryan:2016whc} at the $95\%$ confidence level,
where the SM Higgs production cross sections are assumed. The $ZH$ associated production, in which the  $Z$ boson decays to a charged lepton 
pair, either electron or muon, provides significant constraints to $\mathcal{BR}_{H\to\mathrm{inv}}$ on its own~\cite{Aad:2015pla, Khachatryan:2016whc}. 
The signature is characterized by a large missing transverse momentum recoiling against a charged lepton pair that reconstructs  the $Z$ boson mass. 
The dominant background after signal extraction is $ZZ$~\cite{Choudhury:1993hv, Khachatryan:2016whc, Aaboud:2017bja,Goncalves:2016bkl}, 
where one of the $Z$ bosons decays to a charged lepton pair and the other decays to neutrinos, hence it is an irreducible background. Yet, 
it is possible to measure the polarization of the $Z$ boson from angular distribution of the charged lepton pair for both  the $ZH$ signal and the $ZZ$ background. 

In this paper, we study in detail the possibility of enhancing the $ZH$ signal significance by making use of the difference in $Z$ boson 
polarization between the signal and the dominant $ZZ$ background, following the approach presented in Ref.~\cite{Goncalves:2018fvn}.  
Although this information is disregarded in the present experimental analyses~\cite{Khachatryan:2016whc, Aaboud:2017bja}, we show
that the proposed method can be a key ingredient to pin down the ${H\rightarrow}$~invisibles signal in the $Z(\ell\ell)H$ channel, separating 
the signal  and background underlying  production dynamics more accurately.

The paper is organized as follows. In Section~\ref{sec:polarization}, we show that the signal and the $ZZ$ background predict 
different states of $Z$ polarization and how we use this information for our purpose. In Section~\ref{sec:leptonkinematics}, 
the effects of $Z$ polarization on observables that are constructed with the charged leptons are discussed. 
In Section~\ref{sec:results}, we present the results of our analyses.  In Section~\ref{sec:summary}, we conclude.

%%%%%%%%%%%%%%%%%%%%%%%%%%%%%%%%%%%%%%%%%%%%%%%%%%
\section{$Z$ boson polarization}
\label{sec:polarization}
In general, the $Z\to \ell^+\ell^-$ decay angular distribution for the  ${pp\rightarrow Z(\ell^-_{}\ell^+_{})+X}$  
process can be described  as
\begin{align}
& \frac{1}{\sigma}\frac{d\sigma}{d\cos{\theta}d\phi} =\nonumber \\
 & 1+\cos^2_{}{\theta} 
+ A_{2}^{} (1-3\cos^2_{}{\theta} )
+ A_{3}^{} \sin{2\theta} \cos{\phi}    \nonumber  \\
& + A_{4}^{} \sin^2_{}{\theta} \cos{2\phi} 
+ A_{5}^{} \cos{\theta}
+ A_{6}^{}  \sin{\theta} \cos{\phi}   \nonumber \\
& + A_{7}^{} \sin{\theta} \sin{\phi} 
+ A_{8}^{} \sin{2\theta} \sin{\phi} 
+ A_{9}^{} \sin^2_{}{\theta} \sin{2\phi}\,,
\label{differential}
\end{align}
where $\theta$ ($0 \le \theta \le \pi$) and $\phi$ ($0 \le \phi \le 2\pi$) are the polar and azimuthal 
angles  of the lepton ($\ell^-_{}$) in the $Z$ boson rest frame. We follow the notation of Ref.~\cite{Goncalves:2018fvn}.
The eight coefficients $A_i^{}$ ($i=2$~to~$9$) correspond, in the most general case,  to the number of 
degrees of freedom for the polarization of a spin 1 particle and  uniquely parametrize $Z$ boson 
polarization. In this context, the lepton angular distribution works as an analyser, probing the underlying
production dynamics encoded in the $A_i$ coefficients.
  
%-----------------
 \begin{table*}[th!] 
 \vspace{0.3cm}
 \begin{ruledtabular}
 \begin{tabular}{l*{8}{c}}
       & $A_2^{}$ & $A_3^{}$ & $A_4^{}$ & $A_5^{}$ & $A_6^{}$ & $A_7^{}$ & $A_8^{}$ & $A_9^{}$ \\
\colrule
$ZH$ (LO)    & $0.03(6)$   & $0.2(1)$   & $-80.0(1)$ & $-0.08(8)$ & $-0.01(8)$ & $0.04(8)$ & $0.1(1)$  & $0.1(1)$  \\
$ZH$ (NLO) & $1.7(1)$      & $0.0(3)$  & $-75.0(3)$  & $-0.1(2)$   & $0.6(2)$    & $-0.2(2)$  & $-0.0(3)$ & $0.1(3)$ \\
$ZZ$ (LO)                                 & $48.12(9)$ & $0.3(1)$  & $41.0(1)$   & $0.0(1)$     & $0.2(1)$    & $0.1(1)$   & $-0.1(1)$ & $-0.1(1)$ \\
$ZZ$ (NLO)                              & $49.1(2)$    & $0.0(4)$ & $40.8(4)$    & $-0.3(3)$   & $4.8(3)$    & $0.2(3)$   & $1.0(4)$  & $0.1(4)$ \\
 \end{tabular}
 \end{ruledtabular}
  \caption{Angular coefficients at percent unit, for $Z(\ell^-_{}\ell^+_{})H$ and $Z(\ell^-_{}\ell^+_{})Z$ at  LO and NLO QCD, 
  after the selection in Eq.~\ref{eq:eventcut1}. The statistical uncertainty at the one standard deviation for the last digit is shown in parentheses. 
  \label{table:asymmetries}}
 \end{table*}
%-----------------
    
The value of $A_i$ depends on a reference frame. We choose it following Collins and Soper 
(Collins-Soper frame)~\cite{Collins:1977iv}. This frame is well recognized and the angular coefficients 
for the Drell-Yan $Z$ boson production have been measured by ATLAS~\cite{Aad:2016izn, Aaboud:2017ffb} 
and CMS~\cite{Khachatryan:2015paa}. 

%-----------------
\begin{figure}[b!]
\includegraphics[scale=0.56]{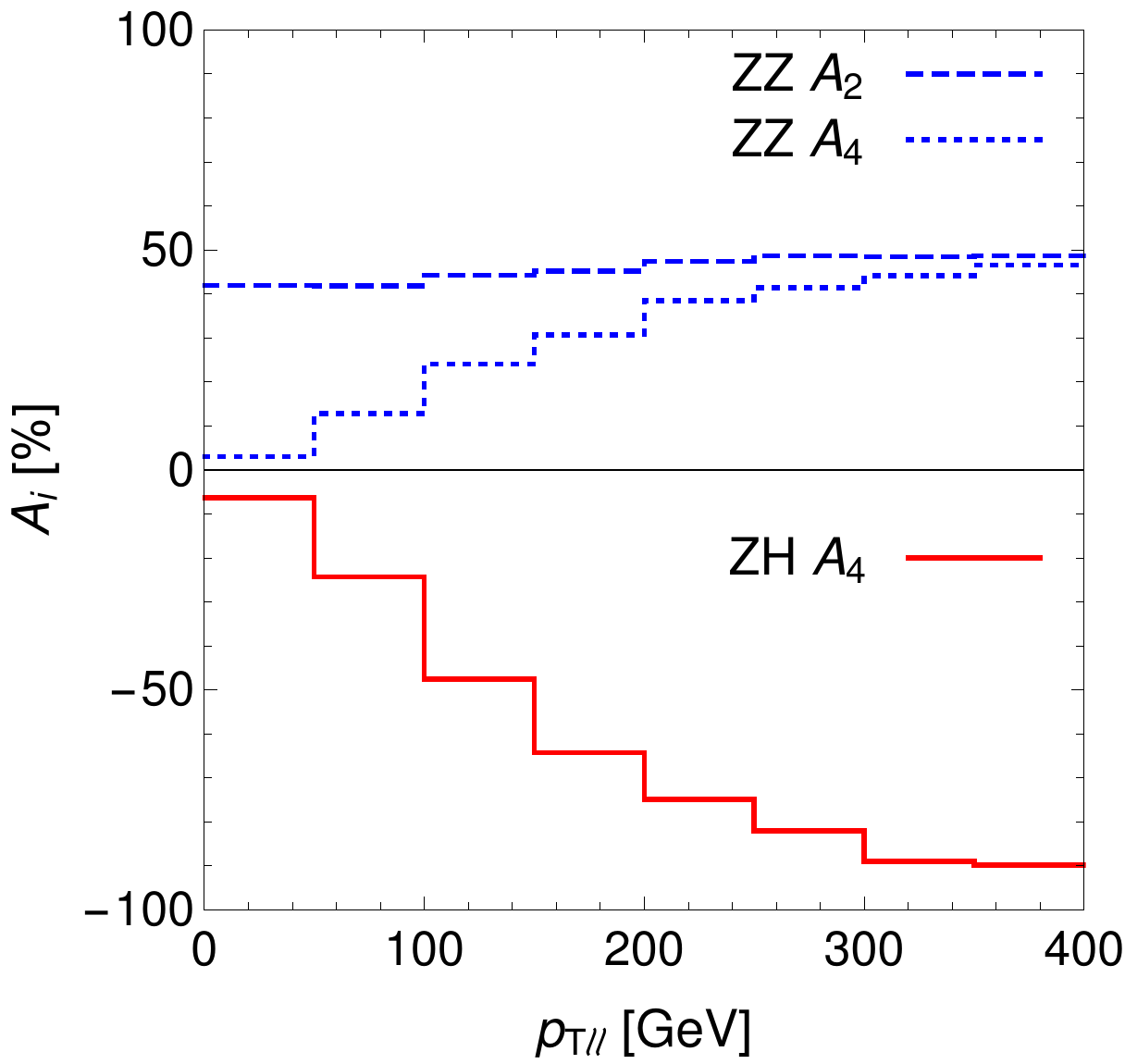}
\caption{Angular coefficients, $A_4^{}$ for $ZH$ (red solid), $A_2^{}$ for $ZZ$ (blue dashed) and $A_4^{}$ for $ZZ$ (blue dotted), calculated in binned 
 $p_{\mathrm{T}\ell\ell}^{}$ regions at the LO accuracy. 
\label{figure:coefficients}}
\end{figure}
%-----------------

To access the potential of a polarization analysis to boost the signal $Z(\ell\ell)H(inv)$ discrimination
against its leading backgrounds, in particular  the dominant  $Z(\ell\ell)Z(\nu\nu)$ contributions, we calculate
the  coefficients $A_i$  at the fixed leading-order (LO) and next-to-leading-order (NLO) QCD with 
MadGraph5\verb|_|aMC@NLO~\cite{Alwall:2014hca} for the 13~TeV LHC, applying the signal selections
\begin{align}
&75~\gev < m_{\ell\ell}^{} < 105~\gev,\ \ p_{\mathrm{T}\ell\ell}^{} > 200~\gev.
 \label{eq:eventcut1}
\end{align}
%
%-----------------
\begin{figure}[th!]
\includegraphics[scale=0.8]{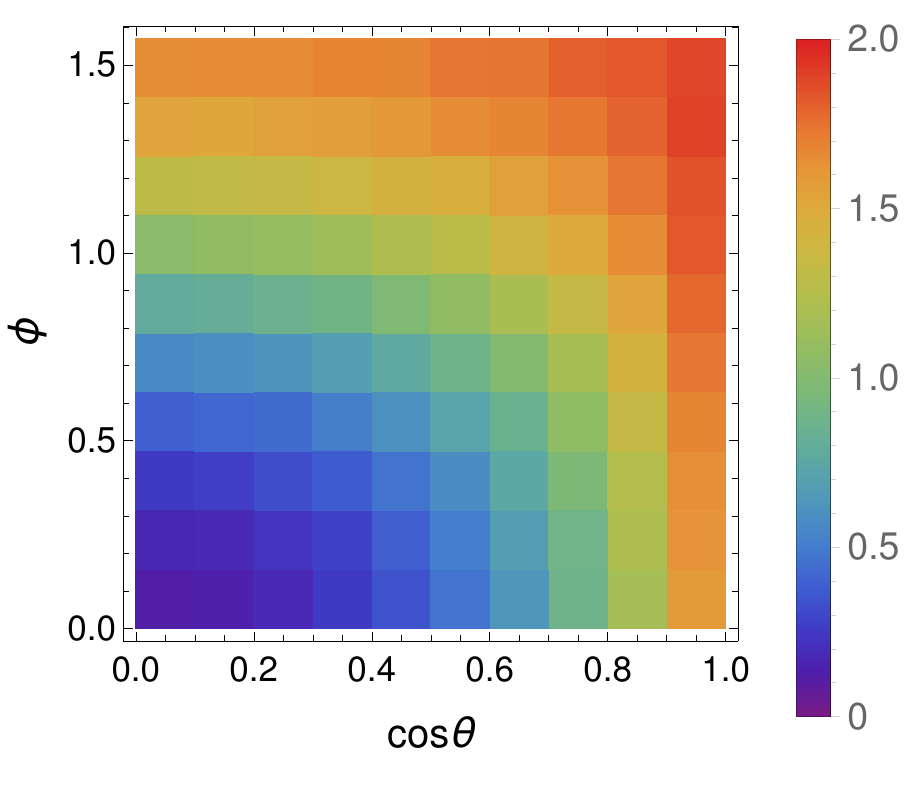}
\caption{Ratio of the normalized $(\cos{\theta},\phi)$ distribution for the $ZH$ process to that for the $ZZ$ background 
process at the LO.
\label{figure:csangles}}
\end{figure}
%-----------------
The $Z(\ell\ell)Z$ background takes into account its interference with $\gamma(\ell\ell)Z$.
The results are summarized, at percent unit,  in Tab.~\ref{table:asymmetries}. 
The $ZH$ signal and $ZZ$ background lepton angular distributions are governed by $A_4^{}$ and $A_{2,4}^{}$, respectively. 
The QCD NLO corrections are visibly large only in $A_{4}^{}$ for $ZH$ and in $A_6^{}$ for $ZZ$. 
The non-zero
$A_6^{}$ in $ZZ$ introduces an asymmetry between $\ell^-_{}$ and $ \ell^+_{}$, which can be 
confirmed by observing the sign  change of the $A_6^{}$ term in Eq.~\ref{differential} after interchanging $\ell^-_{}$ 
and $\ell^+_{}$ ({\it i.e.} $\theta \to \pi - \theta$  {\it and} $\phi \to \phi + \pi$). This indicates that $A_6^{}$ does not contribute 
when we do not distinguish $\ell^-_{}$ and $\ell^+_{}$. We take this approach, in order to make analysis simpler. Because
the difference in $A_6^{}$ between the signal and the background is not large, the loss of  information on $A_6^{}$ 
represents a sub-leading effect.  Consequently, both for the $ZH$ signal and for the $ZZ$ background, Eq.~\ref{differential}
effectively simplifies to 
\begin{align}
& \frac{1}{\sigma}\frac{d\sigma}{d\cos{\theta}d\phi} =\nonumber \\
 & 1+\cos^2_{}{\theta} 
+ A_{2}^{} (1-3\cos^2_{}{\theta} )
 + A_{4}^{} \sin^2_{}{\theta} \cos{2\phi}\;,
\label{differential-2}
\end{align}
%
%-----------------
\begin{figure*}[tbh!]
\centering
\includegraphics[scale=0.45]{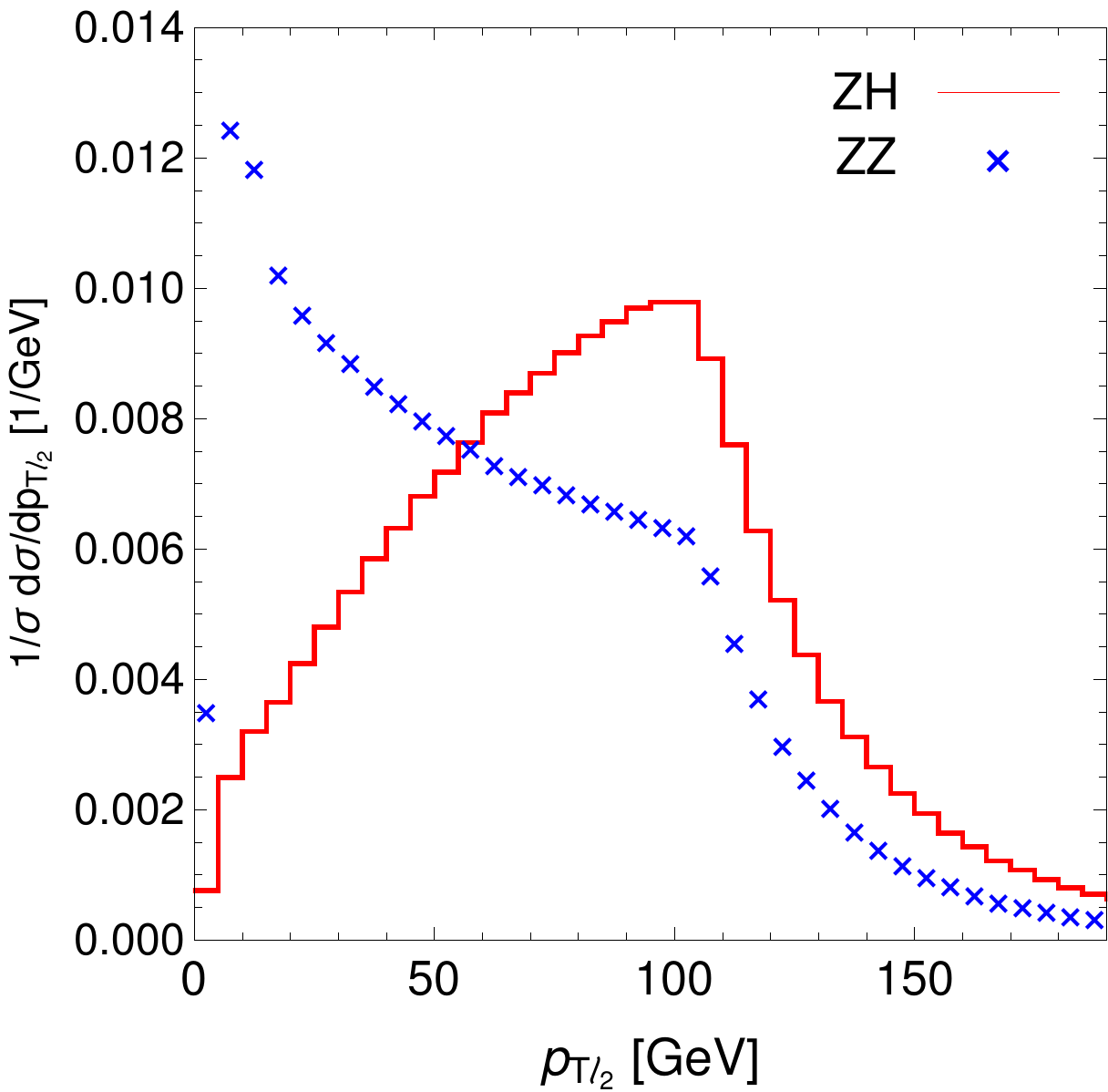}\hspace{.3cm}
\includegraphics[scale=0.44]{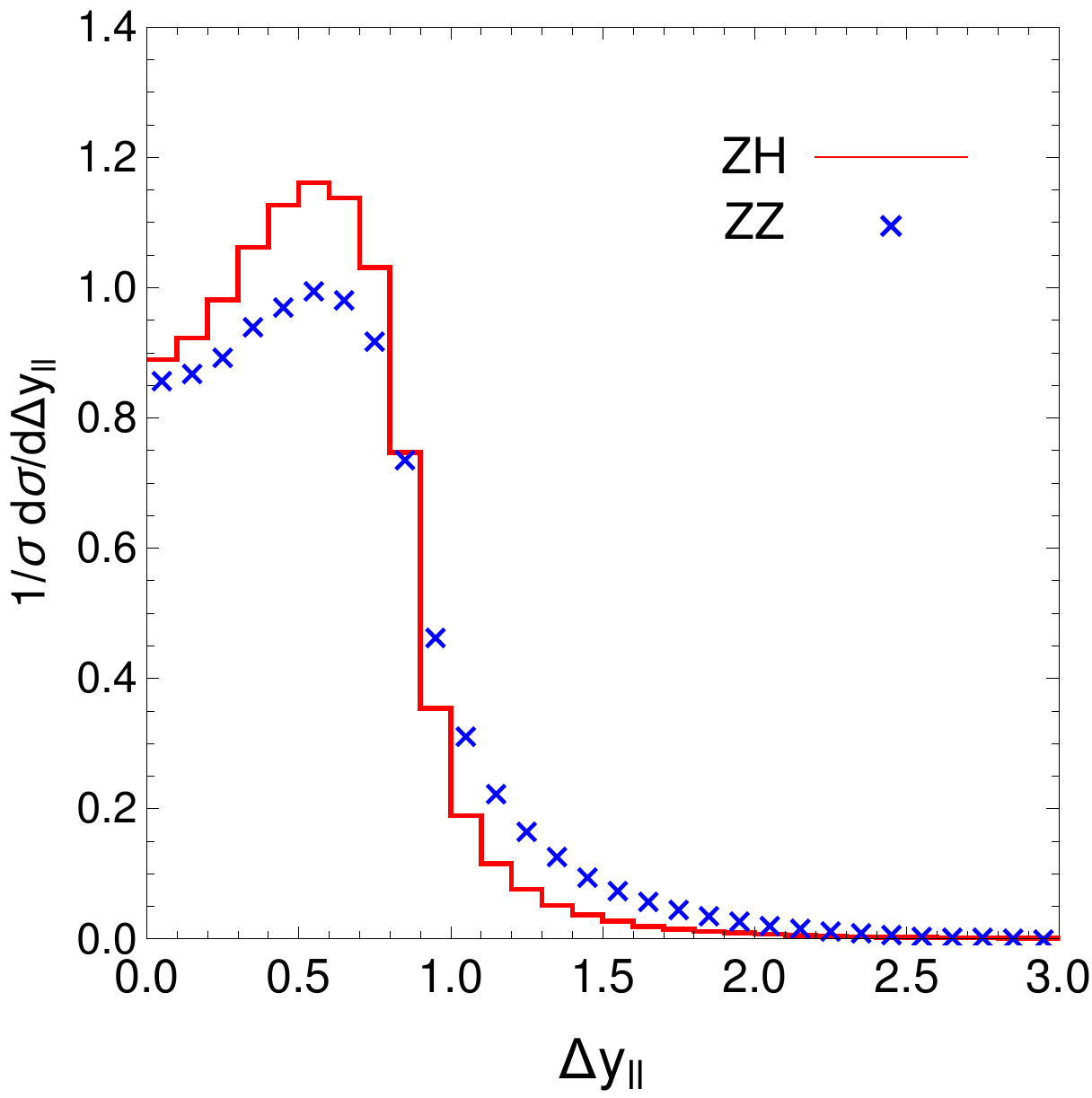}\hspace{.3cm}
\includegraphics[scale=0.43]{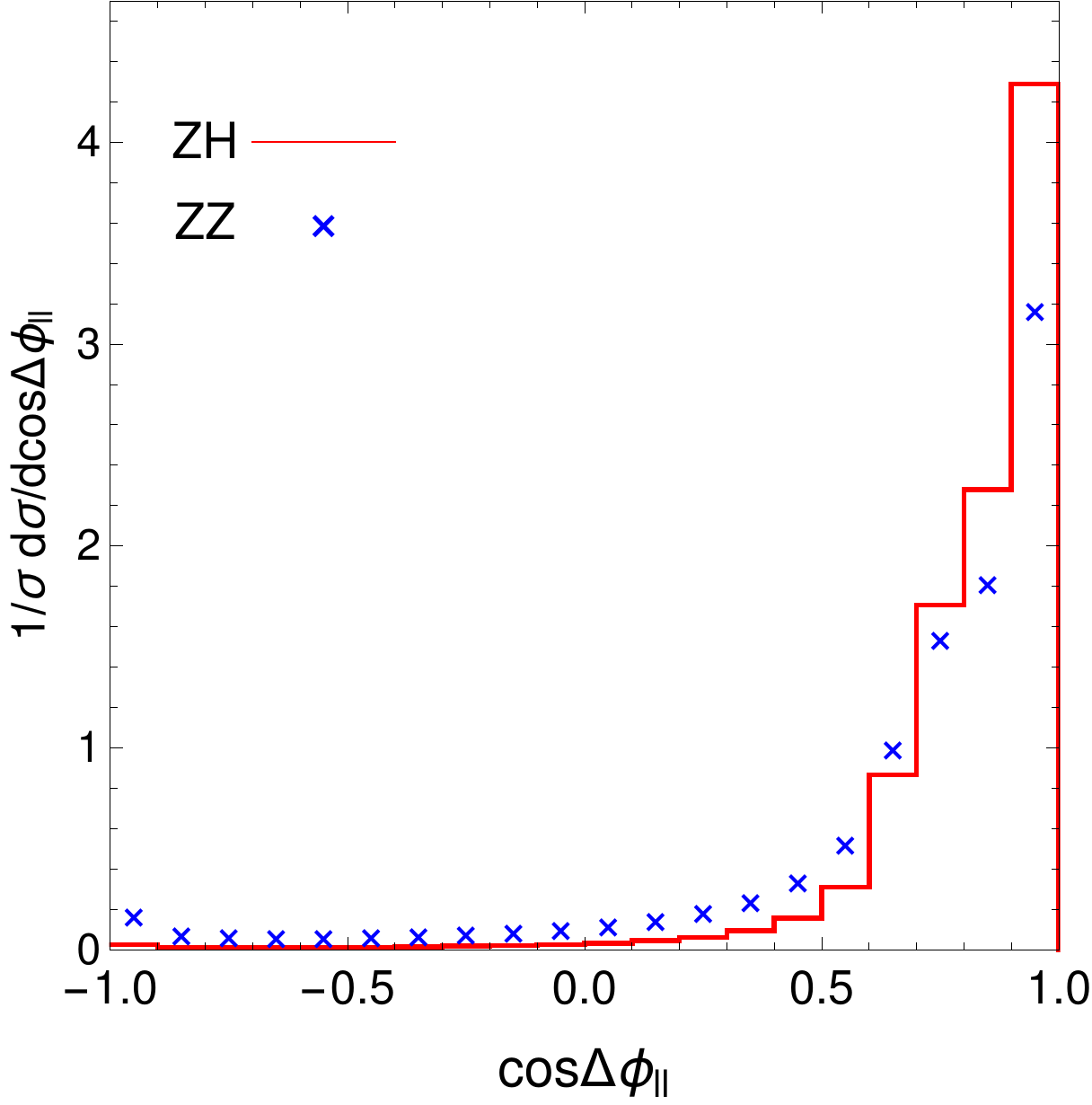}
\caption{\small Normalized distributions of the softer lepton $p_{\mathrm{T}\ell}^{}$ (left), $\Delta y_{\ell\ell}^{}$ 
(middle) and $\cos{ \Delta \phi_{\ell\ell}^{} }$ (right), for$ZH$ (red solid curves) and $ZZ$ (blue $\times$) at the LO,
 imposing the selection in Eq.~\ref{eq:eventcut1}.
\label{figure:leptondist}}
\end{figure*}
%-----------------
where the angles $\theta$ and $\phi$ are defined in the restricted ranges $0 \le \theta \le \pi/2$ and $0 \le \phi \le \pi/2$ 
as a result of not distinguishing $\ell^-_{}$ and $\ell^+_{}$. They can be obtained from
\begin{subequations}
\label{cosphi}
\begin{align}
\cos{\theta} & = \frac{2 \bigl| q^0_{} p_\ell^3 - q^3_{} p_\ell^0 \bigr|}{Q\sqrt{Q^2_{} + |\vec{q}_{\mathrm{T}}^{}|^2_{} }} \,,
 \label{costheta} \\
\cos{\phi} & = \frac{2}{\sin{\theta}}\frac{  \bigl| Q^2_{} \vec{p}_{\mathrm{T}\ell}^{}  \cdot  \vec{q}_{\mathrm{T}}^{} - |\vec{q}_{\mathrm{T}}^{}|^2_{} p_\ell^{} \cdot q \bigr|}{Q^2_{}|\vec{q}_{\mathrm{T}}^{}|\sqrt{Q^2_{} + |\vec{q}_{\mathrm{T}}^{}|^2_{}}} \,, 
\end{align}
\end{subequations}
where $q^{\mu}_{}=(q^0_{}, \vec{q}_{\mathrm{T}}^{}, q^3_{})$ and $p^{\mu}_{\ell}=(p^0_\ell, \vec{p}_{\mathrm{T}\ell }^{}, p^3_\ell)$ 
are four-momenta of the reconstructed $Z$ boson and either of the leptons, respectively, in the laboratory frame and $Q$
is the reconstructed $Z$ invariant mass ($Q=m_{\ell\ell}^{}$)~\cite{Goncalves:2018fvn}. 

In Fig.~\ref{figure:coefficients}, we show the $p_{\mathrm{T}\ell\ell}^{}$ distributions for the  coefficients in Eq.~\ref{differential-2}
calculated at LO, imposing the invariant mass cut in Eq.~\ref{eq:eventcut1}. 
The difference in $A_4^{}$ between $ZH$ and $ZZ$ increases as the $p_{T\ell\ell}$ grows up.
Hence, the  signal and background $Z\to \ell^+\ell^-$ angular distributions  become  more distinct at the boosted regime, 
where they acquire, in particular,  an extra characteristic $\phi$ modulation. In this way, the polarization study dovetails nicely  with the usual boosted 
strategy for the $H\to$~invisibles search in the $Z(\ell\ell)H$ channel.

In Fig.~\ref{figure:csangles}, we show the ratio of the normalized $(\cos{\theta},\phi)$ distribution for the $ZH$ process to that for the $ZZ$ 
process  at the LO, imposing the selections in Eq.~\ref{eq:eventcut1}. The large differences in $A_2^{}$ and $A_4^{}$ between signal and 
background result in phenomenologically relevant kinematic profiles in the two-dimensional $(\cos\theta$, $\phi)$ distribution, where the
signal to background ratio is sensibly enhanced for $(\cos\theta \to 1,\phi\to\pi/2)$ and suppressed for $(\cos\theta \to 0,\phi\to 0)$.
In  Sec.~\ref{sec:results}, we will show that this distribution can be a key element in boosting the signal from background discrimination. 

%-------
\begin{figure*}[t!]
\centering
\includegraphics[scale=0.35]{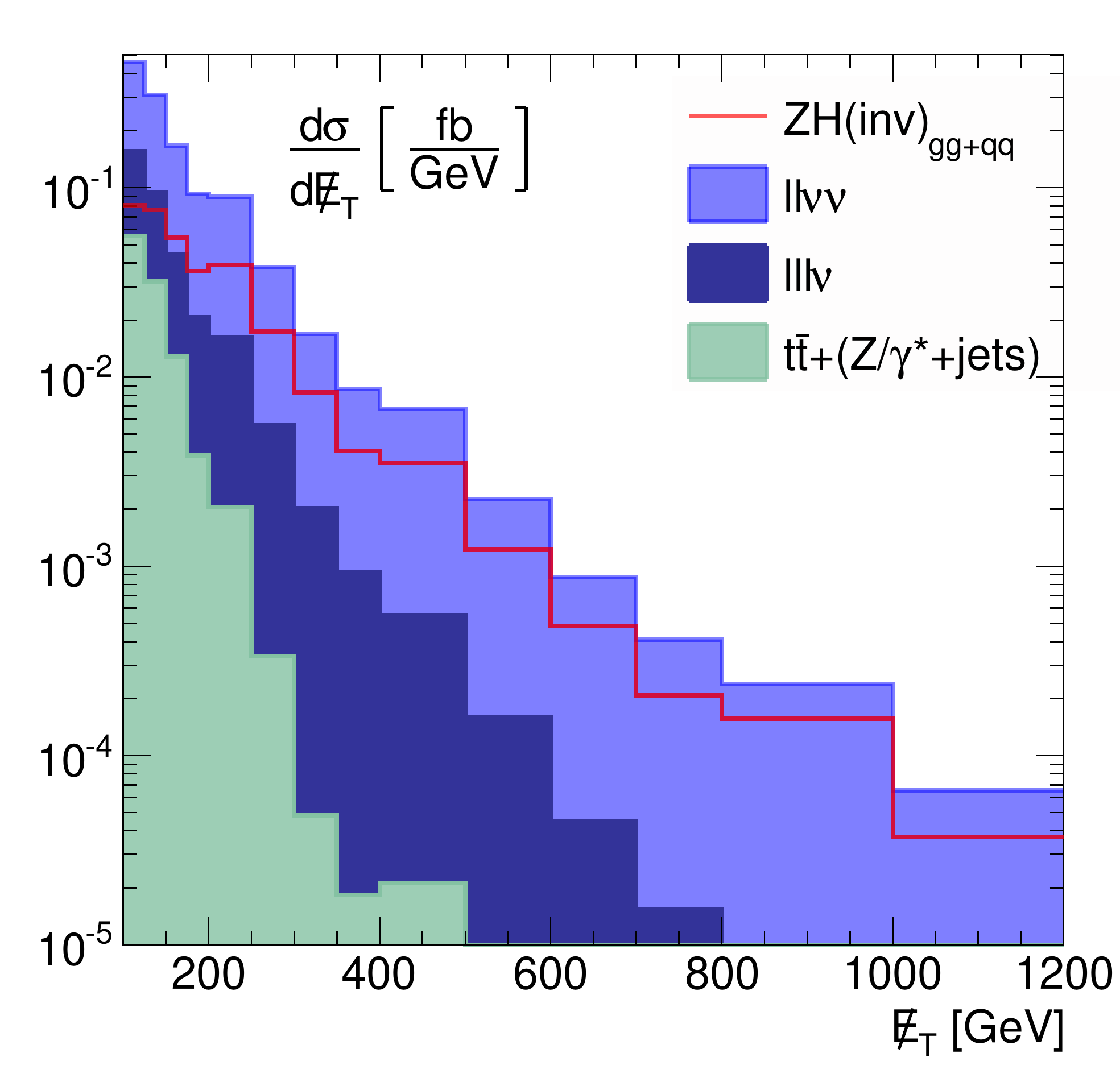}\hspace{1.4cm}
\includegraphics[scale=0.35]{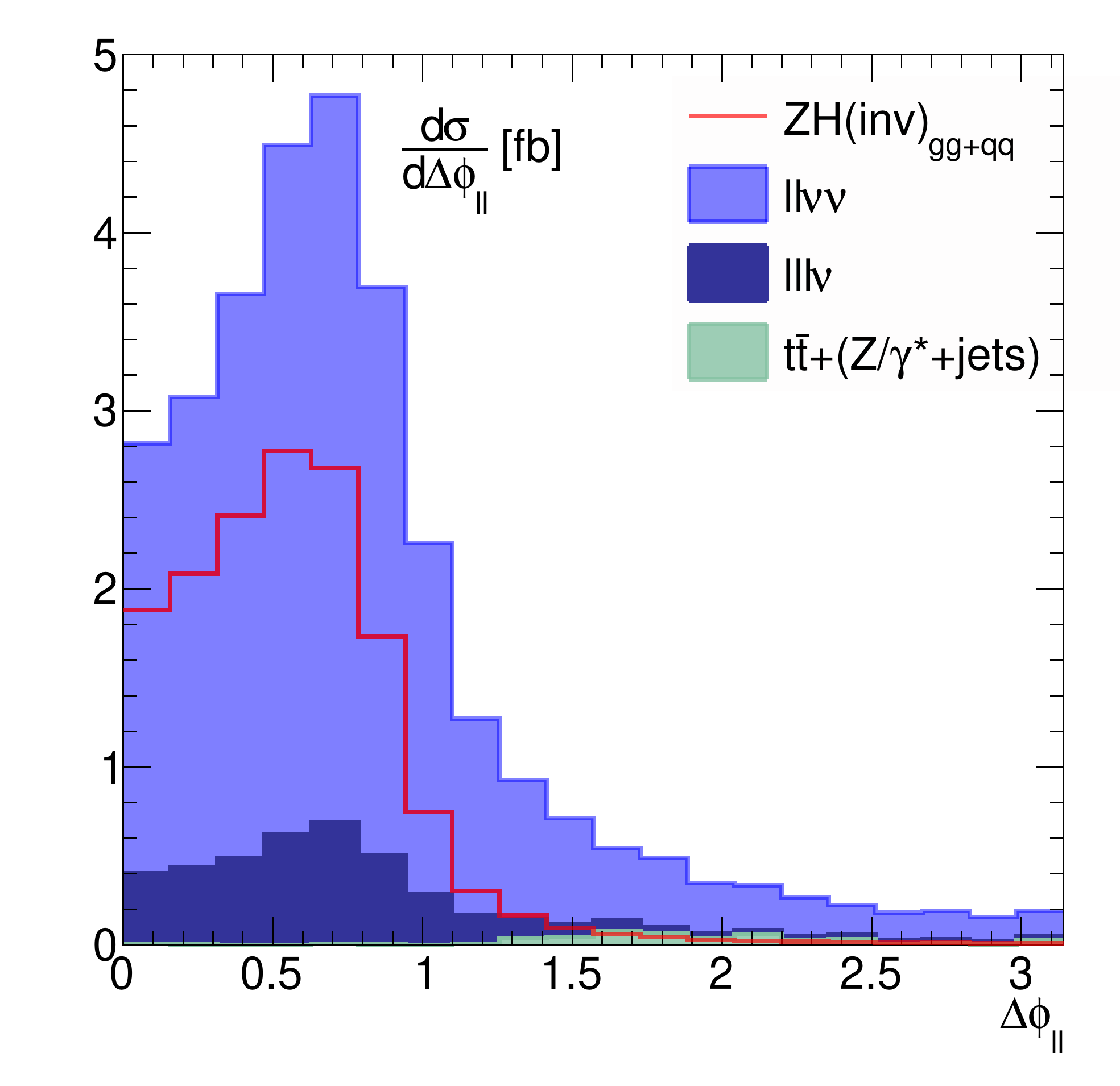}
\caption{Signal (red) and background (blue/green)  $\slashed{E}_T$ (left) and $\Delta\phi_{\ell\ell}$ (right)  distributions.
The background (signal) histograms are (non-)stacked. We apply all the  the selections described in the text but $\slashed{E}_T>200$~GeV
and $\Delta\phi_{\ell\ell}<\pi/2$ (left panel) and $\Delta\phi_{\ell\ell}<\pi/2$ (right panel).  We consider the  $\sqrt{s}=13$~TeV LHC. 
\label{figure:inv}}
\end{figure*}
%-------

%%%%%%%%%%%%%%%%%%%%%%%%%%%%%%%%%%%%%%%%%%%%%%%%%%
\section{Effects of $Z$ polarization on lepton observables}\label{sec:leptonkinematics}
Observables that are constructed by the leptons from the $Z$ boson decay can be, in general, largely
affected by the $Z$ boson polarization. 
Here we illustrate it
with three phenomenologically relevant  observables: the transverse lepton momentum $p_{\mathrm{T}\ell}$, 
the rapidity separation $\Delta y_{\ell\ell}^{}$ $(\ge 0)$ and the azimuthal angle separation $\Delta \phi_{\ell\ell}^{}$  ($0 \le \Delta \phi_{\ell\ell}^{} \le \pi$), all
in the laboratory frame. These observables are used in the signal selection of the ATLAS and CMS analyses~\cite{Aaboud:2017bja,Khachatryan:2016whc}~\footnote{More precisely, $\Delta R_{\ell\ell}^2 = \Delta \phi_{\ell\ell}^2 + \Delta y_{\ell\ell}^2$ is used in the ATLAS analysis.}.   
The transverse momentum $p_{\mathrm{T}}^{}$  of the harder lepton ($\ell_{1}^{}$) and the softer lepton $(\ell_2^{})$ are 
written in terms of the angles $\theta$  and $\phi$ defined in the Collins-Soper frame as~\cite{Goncalves:2018fvn}
\begin{widetext}
\begin{subequations}
\begin{align}
p_{\mathrm{T\ell_{1(2)}}}^{} = \frac{1}{2} \sqrt{ q_{\mathrm{T}}^2 + Q^2_{} \sin^2{\theta} + q_{\mathrm{T}}^2 \sin^2{\theta} \cos^2{\phi} \pm 2 q_{\mathrm{T}}^{} \sqrt{ Q^2_{} + q_{\mathrm{T}}^2 } \sin{\theta} \cos{\phi}  }\;,
 \label{eq:leptonpTmag}
\end{align} 
where $q_{\mathrm{T}}^{}$ is the $Z$ transverse momentum ($q_{\mathrm{T}}^{} = |\vec{q}_{\mathrm{T}}^{}|$). 
In the same manner, $\Delta \phi_{\ell\ell}^{}$ and $\Delta y_{\ell\ell}^{}$ are given by
\begin{align}
e^{2\Delta y_{\ell\ell}^{}}_{} & = \frac{\bigl(\sqrt{ q_{\mathrm{T}}^2 + Q^2_{} } + Q \cos{\theta} \bigr)^2_{} - q_{\mathrm{T}}^{2}\sin^2_{}{\theta} \cos^2_{}{\phi}  }{\bigl(\sqrt{ q_{\mathrm{T}}^2 + Q^2_{} } - Q \cos{\theta} \bigr)^2_{} - q_{\mathrm{T}}^{2}\sin^2_{}{\theta} \cos^2_{}{\phi}}, \\
\cos{ \Delta \phi_{\ell\ell}^{} } 
 & = \frac{ q_{\mathrm{T}}^2 ( 1- \sin^2_{}{\theta} \cos^2_{}{\phi} ) - Q^2_{} \sin^2_{}{\theta} }{ \bigl|q_{\mathrm{T}}^2 ( 1- \sin^2_{}{\theta} \cos^2_{}{\phi} ) - Q^2_{} \sin^2_{}{\theta} \bigr|} \Biggl[ 1 + \frac{4 q_{\mathrm{T}}^2 Q^2_{} \sin^2_{}{\theta} \sin^2_{}{\phi}}{ \bigl\{q_{\mathrm{T}}^2 ( 1- \sin^2_{}{\theta} \cos^2_{}{\phi} ) - Q^2_{} \sin^2_{}{\theta} \bigr\}^2_{} } \Biggr]^{-\frac{1}{2}}_{}. \label{eq:leptondeltaphi} 
\end{align}
\end{subequations}
\end{widetext}
Their $(\theta,\phi)$ dependence in fact shows that these observables are sensitive to the $Z$ polarization. 
Searches for $H\to$~invisibles with the $ZH$ production at the LHC are performed in high $q_{\mathrm{T}}^{}$ 
boosted regions~\cite{Choudhury:1993hv, Khachatryan:2016whc, Aaboud:2017bja}. 
In boosted regions $Q/q_{\mathrm{T}}^{} < 1$, the above observables can be expanded as
\begin{subequations}
\begin{align}
p_{\mathrm{T\ell_{1(2)}}}^{} & = \frac{1}{2} q_{\mathrm{T}}^{} \bigl( 1 \pm \sin{\theta} \cos{\phi} + \mathcal{O}\bigl(Q^2_{}/q_{\mathrm{T}}^2\bigr) \bigr),\label{eq:leptonptlimit} \\
e^{2\Delta y_{\ell\ell}^{}}_{} & = 1 + \frac{4\cos\theta}{1-\sin^2_{}\theta \cos^2_{}\phi} \frac{Q}{q_{\mathrm{T}}^{}} + \mathcal{O}\bigl(Q^2_{}/q_{\mathrm{T}}^2\bigr), \\
\cos{ \Delta \phi_{\ell\ell}^{} } 
 & = 1 + \mathcal{O}\bigl(Q^2_{}/q_{\mathrm{T}}^2\bigr). \label{eq:leptonphi}
\end{align} 
\end{subequations}
The $(\theta,\phi)$ dependence vanishes in $\Delta y_{\ell\ell}^{}$ and $\Delta \phi_{\ell\ell}^{}$ in the limit 
$Q/q_{\mathrm{T}}^{} \to 0$, while it still survives in $p_{\mathrm{T\ell_{1(2)}}}^{}$. 
Therefore, among these three observables, only $p_{\mathrm{T\ell_{1(2)}}}^{}$  is sensitive to $Z$ 
polarization for  highly boosted events. This can be confirmed in Fig.~\ref{figure:leptondist}, in which we show
the normalized distributions for $p_{\mathrm{T}\ell_2^{}}^{}$ (left), $\Delta y_{\ell\ell}^{}$ (middle) and 
$\cos{ \Delta \phi_{\ell\ell}^{} }$ (right), for $ZH$ and $ZZ$ at the LO, imposing the  selections from Eq.~\ref{eq:eventcut1}.  
We observe a large difference only in the $p_{\mathrm{T}\ell_2^{}}^{}$ 
distribution, which originates from the sizable difference in $Z$ polarization shown in  Tab.~\ref{table:asymmetries} 
or in Figs.~\ref{figure:coefficients} and~\ref{figure:csangles}. To summarize,  in high $q_{\mathrm{T}}^{}$ regions, only limited observables  can 
be sensitive to $Z$ polarization and the $p_{\mathrm{T\ell}}^{}$ is one of them.

In Ref.~\cite{Goncalves:2018fvn}, it is found that a higher lepton $p_{\mathrm{T}}^{}$ cut can improve the signal significance as we may expect from  the $p_{\mathrm{T}\ell_2^{}}^{}$ distribution in Fig.~\ref{figure:leptondist}. However, it is also found that the highest signal significance can be achieved by setting the lepton $p_{\mathrm{T}}^{}$ cut as small as possible and analyzing the $(\cos\theta,\phi)$ distribution directly. For our polarization analysis 
based on the $(\cos\theta,\phi)$ distribution, it is  best  to soften the lepton selections as they can generally disturb the $(\cos\theta,\phi)$ 
distribution~\cite{Mirkes:1994eb}. Thus, in our hadron level study presented in Sec.~\ref{sec:results},  we lower the lepton transverse momentum
selection to $p_{\mathrm{T}\ell}^{} > 5$~GeV. 

In the next section,  we will quantify the impact of the $Z$ polarization  to the $Z(\ell\ell)H(inv)$ analysis via a realistic Monte Carlo study. Instead of probing the $Z$ polarization indirectly via the observables $\Delta\phi_{\ell\ell}$, $\Delta y_{\ell\ell}$, or $p_{T\ell_{1(2)}}$,  we will directly explore the lepton angular distribution $(\cos\theta,\phi)$.  

%%%%%%%%%%%%%%%%%%%%%%%%%%%%%%%%%%%%%%%%%%%%%%%%%%
\section{Results}
\label{sec:results}

We now scrutinise the potential improvements from the polarization study  to $pp\rightarrow Z(\ell\ell)H(inv)$ 
analysis. This search is characterised by a boosted leptonic $Z$ boson decay, recoiling against  large transverse 
missing energy from  $H\rightarrow\text{invisibles}$~\cite{Choudhury:1993hv, Khachatryan:2016whc, Aaboud:2017bja,
Goncalves:2016bkl}. The dominant backgrounds for this signature are $t\bar{t}$+jets, $Z/\gamma^*$+jets, and  
diboson pairs $(ZZ, WW)\rightarrow \ell\ell\nu\nu$ and ${ZW\rightarrow \ell\ell\ell\nu}$.

Our signal and background samples are simulated with {\sc Sherpa+OpenLoops}~\cite{Gleisberg:2008ta,Cascioli:2011va,
Denner:2016kdg}. The  $ZH_{DY}$ Drell-Yan signal component,  $t\bar{t}$  and diboson pair samples are generated with the 
MC@NLO algorithm~~\cite{Frixione:2002ik,Hoeche:2011fd}, and the $Z/\gamma^*$+jets is generated up to two extra jet 
emissions at NLO with the MEPS@NLO algorithm~\cite{Hoeche:2012yf}. We also account for the loop-induced gluon fusion 
$ZH_{GF}$ signal component  at leading order accuracy merged up to one extra jet emission via the CKKW 
algorithm~\cite{Goncalves:2015mfa,Goncalves:2016bkl,Catani:2001cc, Hoeche:2009rj}. Spin correlations and finite width effects from  vector bosons are
accounted for  in  our  simulation. Hadronization and underlying event effects are simulated~\cite{Winter:2003tt}.

We start the analysis requiring two same-flavour opposite sign leptons $(\ell=e, \mu)$ with $p_{\mathrm{T}\ell}^{}>5$~GeV and 
$|\eta_\ell|<2.5$, within the $Z$-boson invariant mass window ${|m_{\ell\ell}^{}-m_Z^{}|<15}$~GeV. Events with extra leptons 
are vetoed. Since most of the signal sensitivity resides in the boosted kinematics, we require ${\slashed{E}_\mathrm{T}^{}>200}$~GeV. 

Jets are defined with the anti-$k_\mathrm{T}^{}$  jet algorithm with radius $R=0.4$~\cite{Cacciari:2011ma}, $p_{\mathrm{T}j}^{}>30$~GeV, and 
${|\eta_j^{}|<5}$. To tame the $t\bar{t}$+jets background, we veto events with two or more jets or containing a $b$-jet. Our 
study assumes 70\% $b$-tagging efficiency and 1\% miss-tag rate. We further optimise the signal selection,  requiring  
${\Delta\phi(\ell\ell,\vec{p}_\mathrm{T}^{miss})>2.8}$,  ${|\slashed{E}_\mathrm{T}^{}-p_{\mathrm{T},\ell\ell}^{}|/p_{\mathrm{T},\ell\ell}^{} <0.4}$, transverse mass 
${m_\mathrm{T}^{}>200}$~GeV and $\Delta\phi_{\ell\ell}<\pi/2$, following the CMS analysis~\cite{Khachatryan:2016whc}. An additional selection ${\Delta \phi (\vec{p}_T^{miss},j)>0.5}$
is implemented to the one-jet category  to further suppress the $Z/\gamma^*$+jets background~\cite{Khachatryan:2016whc}.  The resulting signal and 
background $\slashed{E}_T$ distributions   are displayed in Fig.~\ref{figure:inv} (left). The $t\bar{t}$ and $Z/\gamma^*$+jets 
backgrounds get rapidly depleted for large $\slashed{E}_T$, and the diboson contributions ${(ZZ, WW)\rightarrow \ell\ell\nu\nu}$ 
and ${ZW\rightarrow \ell\ell\ell\nu}$ result as the leading background components.

%-------
\begin{figure}[t!]
\centering
\includegraphics[scale=0.59]{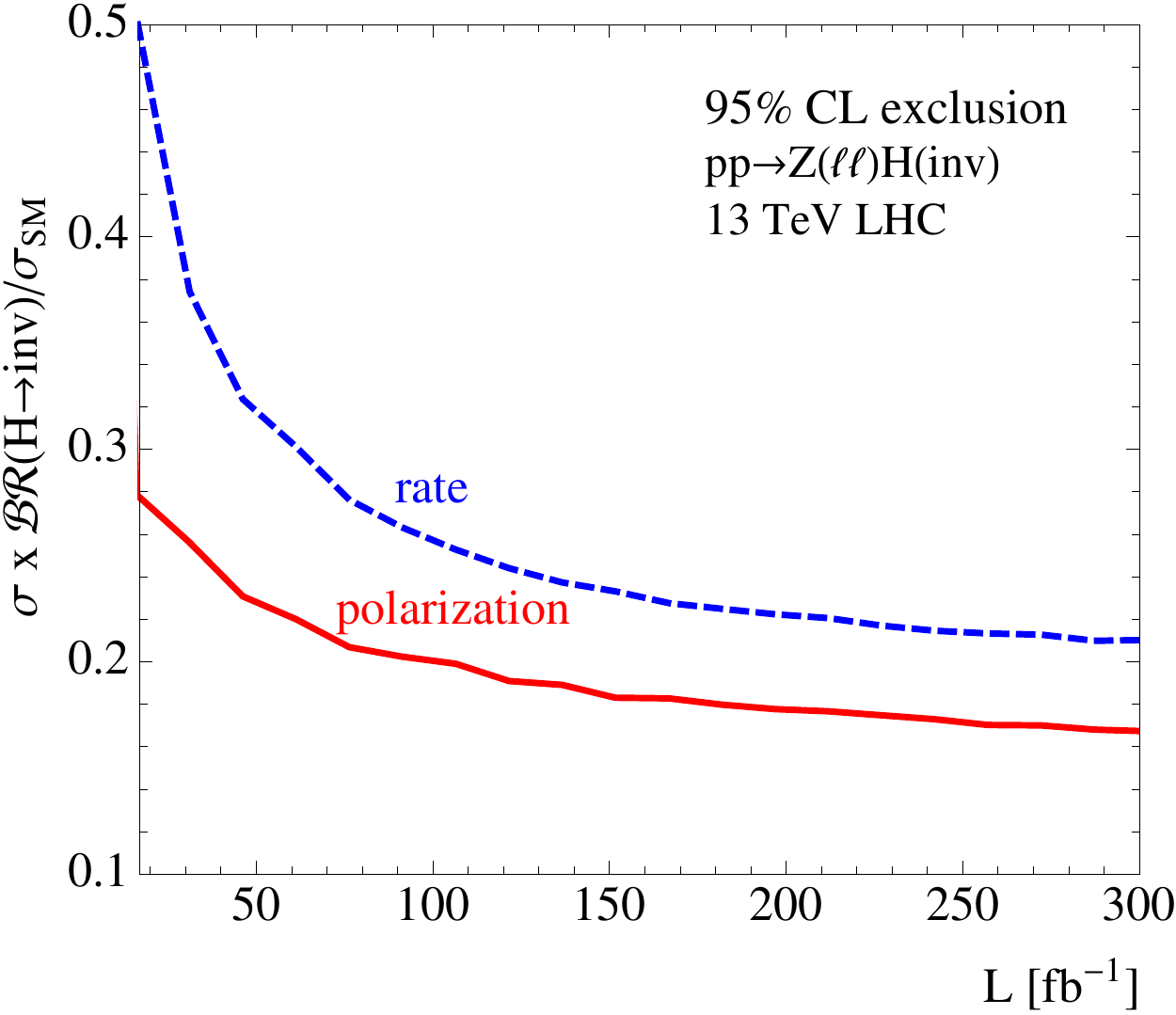}
\caption{Expected 95\% CL upper bound on $\sigma \times \mathcal{BR}(H\rightarrow \text{inv})/\sigma_{SM}$, as a function
of the 13~TeV LHC luminosity, based on the  rate analysis (blue) and on our polarization analysis (red).
\label{figure:bound_inv}}
\end{figure}
%-------

While the selection~$\Delta\phi_{\ell\ell}<\pi/2$ can further suppress some of the backgrounds, 
such as  $t\bar{t}$, see Fig.~\ref{figure:inv} (right), it will have reduced impact in the dominant background 
process $Z(\ell\ell)Z(\nu\nu)$ at the boosted regime. The potential sensitivity in the $\Delta\phi_{\ell\ell}$ 
observable to separate the signal $Z(\ell\ell)H(inv)$ from the background $Z(\ell\ell)Z(\nu\nu)$ channels could arise only from the
$Z$ polarization, however this information is suppressed at the boosted kinematics, as discussed in Sec~\ref{sec:leptonkinematics}. 
Conversely, the direct $(\cos\theta,\phi)$ analysis becomes even more powerful for large transverse momentum, see Fig.~\ref{figure:coefficients},
hence this two-dimensional profile become a key ingredient to achieve more accurate  limits. 

To quantify the possible gains with the polarization study, we perform a binned log-likelihood analysis, invoking the  
CL$_s$ method~\cite{Read:2002hq} on the rate and compare with the  analysis based on the $(\cos\theta,\phi$) distribution.
Our results assume 5\% systematic uncertainty on the background rate modeled as a nuisance parameter. In Fig.~\ref{figure:bound_inv}, 
we show the 95\% confidence level  bound on the $Z(\ell\ell)H$ production times the invisible Higgs branching ratio 
$\mathcal{BR}(H\rightarrow inv)$  normalized  by the SM $Z(\ell\ell)H$ production rate, $\sigma_{SM}$. 
The polarization study largely improves the $H\rightarrow$~invisibles bound and makes it less systematic limited at large collider luminosities.
This is because of the larger signal over background ratio $\mathcal{S}/\mathcal{B}$ for a sizeable portion of the $(\cos \theta,\phi)$ parameter space.
As a result,  we can improve the bound from $\mathcal{BR}(H\rightarrow inv)\lesssim21\%$ to $\lesssim17\%$ by adding the polarization analysis, 
assuming $\mathcal{L}=300$~fb$^{-1}$.

%%%%%%%%%%%%%%%%%%%%%%%%%%%%%%%%%%%%%%%%%%%%%%%%%%
\section{Conclusion}
\label{sec:summary}

In this publication, we present a method to enhance the sensitivity on the $H \to $~invisibles searches with $Z(\ell\ell)H$ 
associated production at the LHC. The proposal relies on the accurate study of the $Z$ boson polarization to disentangle,
with greater precision, the signal and background underlying production dynamics. We first  calculate the complete set of angular 
coefficients  $A_i$  in the $Z$ boson Collins-Soper frame at NLO QCD precision. The signal and background present very distinct
coefficients, consequently, their $Z\to \ell\ell$ angular distributions display relevant phenomenologically differences. Performing a realistic Monte Carlo analysis, we show
that these polarization effects can significantly  improve the $H\to$~invisibles bounds. Assuming an integrated luminosity of  $300\ \mathrm{fb}^{-1}$  at  the 13 TeV LHC,
we achieve a Higgs to invisibles limit that is 20\% stronger by including the polarization effects into the analysis. As this proposal relies only on
lepton reconstruction,  it presents small experimental uncertainties and can be promptly included in the  ATLAS and CMS studies.

%%%%%%%%%%%%%%%%%%%%%%%%%%%%%%%%%%%%%%%%%%%%%%%%%%
\begin{acknowledgments}
DG was funded by U.S. National Science Foundation under the grant PHY-1519175. JN appreciates the support from the 
Alexander von Humboldt Foundation.
\end{acknowledgments}

\bibliography{paper}

\end{document}

%% file: declare.tex
\newcommand\one{\leavevmode\hbox{\small1\normalsize\kern-.33em1}}

\newcommand{\gev}{{\ensuremath\rm GeV}}

\def\slashchar#1{\setbox0=\hbox{$#1$}           % set a box for #1
   \dimen0=\wd0                                 % and get its size
   \setbox1=\hbox{/} \dimen1=\wd1               % get size of /
   \ifdim\dimen0>\dimen1                        % #1 is bigger
      \rlap{\hbox to \dimen0{\hfil/\hfil}}      % so center / in box
      #1                                        % and print #1
   \else                                        % / is bigger
      \rlap{\hbox to \dimen1{\hfil$#1$\hfil}}   % so center #1
      /                                         % and print /
   \fi}

\newcommand{\be}{\begin{eqnarray*}}
\newcommand{\ee}{\end{eqnarray*}}

\newcommand{\bee}{\begin{eqnarray}}
\newcommand{\eee}{\end{eqnarray}}
\newcommand{\beeq}{\begin{equation}}
\newcommand{\eeeq}{\end{equation}}